%% file: DL_for_QGT_arxiv.tex
\documentclass[10pt]{article} 
\usepackage[preprint]{tmlr}

\input{math_commands.tex}

\usepackage{hyperref}
\usepackage{url}
\usepackage{graphicx}
\usepackage{subcaption}
\usepackage{booktabs,caption,siunitx}

\title{Verifiable Deep Quantitative Group Testing}

\author{\name Shreyas Jayant Grampurohit \email shreyasgrampurohit@gmail.com \\
     \addr Department of Electrical Engineering\\
     Indian Institute of Technology (IIT) Bombay, Mumbai, India
     \AND
     \name Satish Mulleti \email mulleti.satish@gmail.com \\
     \addr Department of Electrical Engineering\\
     Indian Institute of Technology (IIT) Bombay, Mumbai, India
     \AND
     \name Ajit Rajwade \email ajitvr@cse.iitb.ac.in\\
     \addr Department of Computer Science and Engineering \\
      Indian Institute of Technology (IIT) Bombay, Mumbai, India
      }

\newcommand{\y}{\boldsymbol{y}}
\newcommand{\x}{\boldsymbol{x}}
\newcommand{\A}{\boldsymbol{A}}

\def\month{MM}  
\def\year{YYYY} 
\def\openreview{\url{https://openreview.net/forum?id=XXXX}} 

\begin{document}

\maketitle

\begin{abstract} 
    We present a neural network-based framework for solving the quantitative group testing (QGT) problem that achieves both high decoding accuracy and structural verifiability. In QGT, the objective is to identify a small subset of defective items among $N$ candidates using only $M \ll N$ pooled tests, each reporting the number of defectives in the tested subset. We train a multi-layer perceptron to map noisy measurement vectors to binary defect indicators, achieving accurate and robust recovery even under sparse, bounded perturbations. Beyond accuracy, we show that the trained network implicitly learns the underlying pooling structure that links items to tests, allowing this structure to be recovered directly from the network’s Jacobian. This indicates that the model does not merely memorize training patterns but internalizes the true combinatorial relationships governing QGT. Our findings reveal that standard feedforward architectures can learn verifiable inverse mappings in structured combinatorial recovery problems.
\end{abstract}

\section{Introduction}
Group testing (GT) is the problem where a set of samples of $N$ subjects, of which only $K \ll N$ are defective/diseased, are tested not individually, but by means of $M$ pool tests, where  $K< M \ll N.$ A pooled test consists of testing a mixture of small, equal-volume portions of the samples of a subset of the $N$ subjects. Typically, each subject participates in more than one pool. If $K \ll N$, group testing theory shows conditions under which accurate or near-accurate recovery of the health status of the $N$ subjects is possible \cite{aldridge2019group,du1999combinatorial}, from the $M$ pooled test results. The advantage is that group testing has the capability of saving on the number of tests and on testing resources, including testing time. Group testing has a long history dating back to the seminal work of Dorfman \cite{dorfman1943detection}, and has seen renewed interest ever since the COVID-19 pandemic \cite{ghosh2021compressed,bharadwaja2022recovery,cohen2021multi}.

In typical group testing, the outcome of each test is binary: a 1 if the pool contains at least one defective sample, or a 0 if the pool contains no defective samples. The defect status of each individual sample is also binary: a 1 for a defective sample and a 0 for a non-defective (healthy) sample. In \emph{quantitative} group testing (QGT), which is the subject of this paper, the outcome of each pooled test is a non-negative integer indicating the \emph{number} of defective items in each pool (which could be from 0 to the number of items in the pool). Given a vector $\y \in \mathbb{Z}^{M}_{+,0}$ of pooled test outcomes, and a binary pooling matrix $\A \in \{0,1\}^{M \times N}$, the aim is to recover the health status $\x \in \{0,1\}^{N}$ of the $N$ participating samples (one per subject). Such quantitative problems arise in diverse applications such as bio-informatics \cite{cao2014quantitative}, communications \cite{wang2015group}, and efficient image moderation \cite{ghosh2023efficient}. Quantitative group testing has also been a topic of research in theoretical computer science \cite{sebHo1985two,bshouty2009optimal}. In some references, QGT is also called ``group testing under sum observations'' \cite{wang2015group} or the ``pooled data problem'' \cite{tan2024approximate,scarlett2017phase}. In the latter case, defective items are considered to be of some $G$ different types, and the pooled results report the number of defectives in each type. Thus, QGT is a special case of the pooled data problem with $G = 1$.  

Broadly, QGT strategies fall into two categories: \textit{non-adaptive}, where all tests are specified a priori and performed in parallel, and \textit{adaptive}, where tests are designed sequentially based on previous outcomes. While adaptive strategies generally require fewer measurements to achieve recovery \cite{li2020combinatorial}, they incur higher latency and implementation complexity due to the sequential nature of the testing process. Recent works have sought to optimize this trade-off; for instance, \cite{soleymani2024tunable} introduced a framework with ``tunable'' adaptation to bridge the gap between the two paradigms, while \cite{li2020combinatorial} provided explicit adaptive constructions robust to adversarial noise. Despite these theoretical advances, non-adaptive schemes remain indispensable for applications requiring massive parallelism and minimal latency, motivating the data-driven approach of this work.

In this work, we train a multi-layer perceptron (MLP) on $n^\text{MLP}_\text{train}$ pairs of the form $\{(\x_i, \y_i)\}_{i=1}^{n^\text{MLP}_\text{train}}$, where $\y_i \in \mathbb{Z}^{M}_{+,0}$ denotes the pooled test outcomes corresponding to a sparse binary signal $\x_i \in \{0,1\}^N$ with $\mathbb{E}[\|\x_i\|_0] = K$. The forward relationship between $\y_i$ and $\x_i$ is governed by the known pooling matrix $\boldsymbol{A}$, as described in \eqref{eq:ideal_qgt}. Although the MLP is trained on only $n^\text{MLP}_\text{train} \ll \binom{N}{K}$ sample pairs of the form $\{(\x_i, \y_i)\}_{i=1}^{n^\text{MLP}_\text{train}}$, we demonstrate that it generalizes effectively to previously unseen supports and remains robust to measurement noise. Beyond achieving accurate recovery, we show that the trained network implicitly learns the latent pooling structure encoded in $\A$, even though $\A$ was not explicitly provided as input during training. This reveals a form of verifiability, suggesting that the model has internalized the true combinatorial relationships governing QGT rather than memorizing dataset-specific correlations. To the best of our knowledge, this is the first application of neural networks to quantitative group testing. 

\section{Related Work}
\label{sec:relatedwork}

Our data-driven, non-adaptive framework is best understood in contrast to existing model-driven algorithms, which are often designed for different noise assumptions and require explicit knowledge of the signal's sparsity.

The theoretical foundations of the \emph{pooled data problem}, of which QGT is a special case, were established by \cite{scarlett2017phase}, who provided sharp information-theoretic characterizations of the fundamental recovery limits under both noiseless and noisy measurement models. Their work demonstrated phase transitions in sample complexity and highlighted how even mild noise can dramatically increase the difficulty of reliable recovery. However, these results focus on asymptotic limits rather than algorithmic realizations, motivating subsequent works toward practical decoders.

A prominent algorithmic baseline for QGT is the Approximate Message Passing (AMP) framework. \cite{tan2024approximate} recently provided the first rigorous performance guarantees for an AMP algorithm in this domain, analyzing its performance under dense statistical noise, where every measurement is assumed to be independently perturbed. While powerful, AMP is fundamentally model-driven; as our own results in Section~\ref{sec:results} confirm, its performance is critically dependent on having an accurate estimate of the sparsity $K$ for each sample. In sharp contrast, our MLP-based decoder is trained on signals that are sparse \emph{in expectation} and does not require knowledge of signal sparsity at inference time. Furthermore, our framework is the first to demonstrate a high-performance decoder for the practical sparse noise model, a scenario distinct from the dense noise models studied in the AMP literature.

Other non-adaptive approaches are based on explicit combinatorial constructions. \cite{karimi2019sparse} proposed a low-complexity decoder using sparse graph codes. \cite{gebhard2022parallel} introduced an efficient, greedy ``Maximum Neighborhood'' algorithm. More recently, \cite{soleymani2024nonadaptive} improved the theoretical decoding complexity for non-adaptive QGT using a ``concatenated construction''. A thread common to these methods is their reliance on a \emph{combinatorial model} in which measurements are assumed to be perfectly noiseless, and the sparsity $K$ is a fixed, known-in-advance parameter. Our framework, by contrast, is designed to be robust to sparse measurement noise and learns from a probabilistic distribution, making it unconstrained by a fixed $K$ and inherently robust to the natural sparsity variations expected in practical applications.

Finally, our work can be contrasted with the theoretical analysis of \cite{li2020combinatorial}, who studied a \emph{dense, worst-case} noise model. Their work is primarily theoretical; while they characterize the fundamental limits for non-adaptive recovery, they explicitly state that an optimal non-adaptive construction ``remains open''. The only optimal-rate algorithm that they fully construct is \emph{adaptive}. Our trained MLP thus serves as a constructive, practical, and non-adaptive decoder for the distinct and practical sparse noise model, providing a concrete algorithmic solution where prior work had identified a theoretical gap.

\section{Problem Formulation}
\label{sec:problemformulation}
We consider the quantitative group testing (QGT) model in which an unknown binary vector $\x \in \{0,1\}^N$ represents the defect status of $N$ items. Each entry $x_j$ is independently drawn from a Bernoulli distribution with
\begin{equation}
\Pr(x_j = 1) = \frac{K}{N}, \quad \Pr(x_j = 0) = 1 - \frac{K}{N},
\label{eq:gen_K}
\end{equation}
where $K \ll N$ denotes the expected number of defectives. Hence, $\x$ is \emph{sparse in expectation}, with $\mathbb{E}[\|\x\|_0] = K.$

Each of the $M$ pooled tests corresponds to a subset of items, defined by the pooling matrix $\A \in \{0,1\}^{M \times N}$. The $(i,j)$-th entry of $\A$ equals 1 if item $j$ is included in pool $i$, and 0 otherwise. In the ideal (noiseless) setting, the QGT measurement model is given by
\begin{equation}
\y = \A\x,
\label{eq:ideal_qgt}
\end{equation}
where $\y \in \mathbb{Z}_+^M$ represents the vector of pooled test outcomes, with each entry $y_i$ indicating the number of defectives in pool $i$.

Throughout this work, the pooling matrix $\A$ is generated by drawing each entry from a Bernoulli distribution with parameter $0.5$. Such random designs 
are standard in the QGT and pooled-data literature (e.g., \cite{scarlett2017phase, tan2024approximate}), 
as they provide well-conditioned measurement ensembles with high probability and avoid structural biases introduced by deterministic constructions. In particular, Bernoulli designs ensure that each item participates in roughly half of the tests, yielding a forward operator that satisfies desirable expansion and incoherence properties. Since our goal is to study the behavior of a learned decoder rather than optimize the pooling design itself, the Bernoulli ensemble serves as a neutral, unbiased, and 
informationally strong choice for all experiments.
In practice, measurement noise or small experimental errors may perturb the test outcomes. We model this using a sparse, integer-valued noise vector $\boldsymbol{\eta} \in \mathbb{Z}^M$. Each entry $\eta_i$ is independently drawn from a mixture distribution: with probability $S/N$, $\eta_i$ takes a value chosen uniformly at random from the set $\{-D, \dots, D\}$; otherwise, $\eta_i = 0$. Here, $D \in \mathbb{Z}_+$ bounds the maximum noise magnitude, and the parameter $S$ controls the expected sparsity of the noise relative to the signal dimension. The resulting noisy measurement model becomes
\begin{equation}
\y = \A\x + \boldsymbol{\eta}.
\label{eq:noisy_model}
\end{equation}

The goal of QGT is to accurately recover the unknown binary vector $\x$, or equivalently, its support set of defectives, from the noisy measurements $\y$ and the known pooling matrix $\A$. Classical decoding methods for QGT rely on combinatorial or optimization-based procedures that may be computationally intensive or sensitive to noise. Instead, we propose a data-driven framework that learns a mapping
\begin{equation}
f_{\boldsymbol{\theta}}: \mathbb{Z}^M \rightarrow \mathbb{R}^N,
\end{equation}
parameterized by neural network weights $\boldsymbol{\theta}$, such that $\hat{\x} = f_{\boldsymbol{\theta}}(\y)$ approximates the true defect indicator vector $\x$ with high accuracy and robustness to sparse integer noise.

\section{Method}

\label{sec:method}
We now describe our proposed framework for quantitative group testing. Given the measurement model introduced in Section \ref{sec:problemformulation},
$
\y = \A\x + \boldsymbol{\eta}, $
where $\A \in \{0,1\}^{M\times N}$ is the known pooling matrix, $\x \in \{0,1\}^N$ is the unknown defect indicator vector, and $\boldsymbol{\eta}$ denotes sparse bounded noise, our goal is to learn a data-driven mapping that recovers $\x$ from $\y$.

The framework consists of two complementary components:
(i) Learning to Decode, which trains a neural model to estimate $\boldsymbol{x}$ from noisy measurements; and
(ii) Structural Verifiability Analysis, which analyzes whether the learned mapping is structurally consistent with the true pooling design $\boldsymbol{A}$.

\subsection{Learning to Decode: MLP-based Recovery}
We treat the recovery of $\boldsymbol{x}$ from $\boldsymbol{y}$ as a supervised learning problem.
A multi-layer perceptron (MLP), denoted $f_{\boldsymbol{\theta}}:\mathbb{Z}^M \to \mathbb{R}^N$, is trained to approximate the mapping between pooled test outcomes and defect indicator vectors. Dataset $\mathcal{D}^\text{MLP}$ $\{(\boldsymbol{y}_i, \boldsymbol{x}_i)\}_{i=1}^{n}$ is generated synthetically according to the probabilistic model described in Section~\ref{sec:problemformulation}, using a fixed pooling design across all samples. $\mathcal{D}^\text{MLP}$ is split into training, validation, and test datasets $\mathcal{D}_\text{train}^\text{MLP}$, $\mathcal{D}_\text{val}^\text{MLP}$, and $\mathcal{D}_\text{test}^\text{MLP}$ of sizes $n_\text{train}^\text{MLP}$, $n_\text{val}^\text{MLP}$, and $n_\text{test}^\text{MLP}$. 

\noindent \textbf{Generalizability of the Data-Driven Framework:} 
It is important to note that while the results presented in this work are derived from the specific probabilistic model detailed in Section \ref{sec:problemformulation}, the proposed framework is inherently distribution-agnostic. Unlike model-driven algorithms that often rely on rigid signal assumptions (e.g., worst-case sparsity), our training methodology allows for the incorporation of \textit{any} known signal prior directly into the dataset generation process. If domain-specific knowledge, such as correlations between defective items or specific non-uniform defect probabilities, is available, it can be seamlessly integrated into the training set $\mathcal{D}^\text{MLP}_\text{train}$. Similarly, precise noise characteristics can be simulated during training, enabling the MLP to learn a specialized inverse mapping tailored to the exact distribution of the deployment environment.

The parameters $\boldsymbol{\theta}$ are optimized using a \emph{balanced mean squared error} loss, which equally averages the mean squared errors computed over defective and non-defective entries of $\boldsymbol{x}$. 
This prevents the optimization from being dominated by the majority of zero entries while maintaining a symmetric quadratic objective. 

For a given set of training pairs $\{(\y_i, \x_i)\}_{i=1}^{n_{train}^{MLP}}$, the loss is defined as: 
\begin{equation}
\mathcal{L}(\boldsymbol{\theta})
= \frac{1}{2} \sum_{i=1}^{n_{train}^{MLP}} \left(
\frac{\sum_{j=1}^{N} (x_{ij} - \hat{x}_{ij})^2 \, \mathbf{1}{\{x_{ij} = 1\}}}{\sum_{j=1}^{N} \mathbf{1}\{x_{ij} = 1\}}
+
\frac{\sum_{j=1}^{N} (x_{ij} - \hat{x}_{ij})^2 \, \mathbf{1}\{x_{ij} = 0\}}{\sum_{j=1}^{N} \mathbf{1}\{x_{ij} = 0\}}
\right),
\label{eq:balanced_mse}
\end{equation}
where $\hat{\boldsymbol{x}} = f_{\boldsymbol{\theta}}(\boldsymbol{y})$ denotes the network output, and $\mathbf{1}\{z\}$ is the indicator function, which outputs 1 if the predicate $z$ is true and outputs 0 otherwise. $x_{ij}$ and $\hat{x}_{ij}$ are the $j$-th elements of $\x_i$ and $\hat{\x}_i$, respectively. We found that this balanced quadratic formulation yields smoother convergence and more stable gradients in the sparse regime, where $K/N \ll 1$.
The network outputs $\hat{\boldsymbol{x}} \in \mathbb{R}^N$ and is trained using the Adam optimizer with early stopping based on validation loss. 
These choices were found to provide stable convergence across training runs. 
    Complete architectural and training details, including hidden-layer configuration, activation functions, and hyperparameters, are reported in Appendix \ref{app:arch_details}. 

During inference, a binary estimate $\hat{\boldsymbol{x}}_{\text{bin}}$ is obtained via elementwise thresholding of $\hat{\x}$:
\begin{align*}
\hat{x}_{\text{bin},j} =
\begin{cases}
1, & \text{if } \hat{x}_j \ge \tau,\\[4pt]
0, & \text{otherwise,}
\end{cases}
\end{align*}
where the threshold $\tau$ is chosen to maximize the Success Rate on a validation set, defined as the fraction of samples for which $\hat{\boldsymbol{x}}_{\text{bin}} = \boldsymbol{x}$.

Empirically, the trained network generalizes well to unseen sparsity patterns and noise configurations, even when the number of training samples $n_\text{train}^\text{MLP}$ is far smaller than $\binom{N}{K}$. Intuitively, however, we require the training set to be sufficiently diverse such that the union of support sets covers all indices $\{1, \dots, N\}$, ensuring that the model has the opportunity to learn the forward mapping for every item. This ability to generalize suggests that the model does not merely memorize defect configurations but learns the underlying structural relationships induced by $\boldsymbol{A}$.

\subsection{Structural Verifiability Analysis}

While the MLP decoder achieves strong empirical recovery performance, it is important to assess whether the learned mapping is consistent with the true measurement process defined by $\boldsymbol{A}$.
To this end, we introduce a structural verifiability analysis, which examines whether small perturbations in the network’s input produce output variations consistent with the known pooling structure.

Intuitively, if the decoder has learned a faithful inverse mapping of the QGT process, its local behavior around any input $\boldsymbol{y}_i$ should approximately invert the forward transformation $\boldsymbol{A}$.
We test this hypothesis by estimating an effective pooling matrix $\boldsymbol{\hat{A}}$ that best aligns the network’s local behavior with the identity transformation across a set of $T$ test samples:
\begin{equation}
\boldsymbol{\hat{A}} =
\arg\min_{\boldsymbol{C} \in \mathbb{R}^{M\times N}}
\sum_{i=1}^{T} \|\boldsymbol{I}_M - \boldsymbol{C}\boldsymbol{B}_i\|_F^2,
\label{eq:verifiability}
\end{equation}
where $\|\cdot\|_F$ denotes the Frobenius norm and $\boldsymbol{B}_i$ represents the local sensitivity (Jacobian) of the network output with respect to its input for the $i$-th test sample. In other words, we have
\begin{equation}
\boldsymbol{B}_i = \dfrac{\partial \hat{\boldsymbol{x}}_i}{\partial \boldsymbol{y}_i}, 
\end{equation}
and $\boldsymbol{B}_i$ is a matrix of size $N \times M$. The optimization in \eqref{eq:verifiability} seeks a matrix $\boldsymbol{C}$, interpreted as an estimate of the pooling matrix $\boldsymbol{A}$, that, when composed with the network’s local behavior, approximates the identity transformation.

Since \eqref{eq:verifiability} is a quadratic least-squares problem, its relaxed solution admits a closed form given by
\begin{equation}
    \boldsymbol{\hat{A}}_{\text{relaxed}} = \left(\sum_{i=1}^{T}\boldsymbol{B}_i^{\top}\right)
    \left(\sum_{i=1}^{T}\boldsymbol{B}_i\boldsymbol{B}_i^{\top}\right)^{-1}.
    \label{eq:arelaxed}
\end{equation}
This $\boldsymbol{\hat{A}}_{\text{relaxed}}$ serves as a continuous-valued estimate of the underlying pooling design, reflecting the aggregate alignment between the network’s learned local mappings and the identity operator.

To obtain a binary pooling matrix, we binarize $\boldsymbol{\hat{A}}_{\text{relaxed}}$ using 
K-means clustering with two clusters. The cluster with the smaller centroid is assigned the value $0$, and the other is assigned the value 1, yielding the final binary estimate $\hat{\boldsymbol{A}}$.

The resulting matrix $\boldsymbol{\hat{A}}$ is then compared with the true pooling design $\boldsymbol{A}$ to quantify structural consistency.
A small difference between the two matrices indicates that the decoder’s learned mapping behaves locally as an approximate inverse of the true forward process, confirming structural verifiability.
Empirically, this analysis reveals that the trained network implicitly reconstructs the original measurement structure despite never being given $\boldsymbol{A}$ during training.

\section{Results and Discussion}
\label{sec:results}

\subsection{Experimental Setup and Comparisons}

We evaluate the proposed MLP-based decoder against the \emph{Approximate Message Passing (AMP)} algorithm from \cite{tan2024approximate}, a strong baseline for sparse recovery in QGT that assumes prior knowledge of signal sparsity. All methods are tested on datasets drawn from the same generative model described in Section~\ref{sec:problemformulation}: each item is defective with a probability $K/N$, the measurement noise vector $\boldsymbol{\eta}$ is sparse, with non-zero entries occurring with a probability $S/N$, where $S < N$, and the absolute value of each non-zero entry in $\boldsymbol{\eta}$ is an integer bounded by $D$. The MLP is trained on samples drawn from the same distribution, ensuring a consistent and fair comparison across all methods.

We consider three AMP variants that differ in their assumptions about sparsity information:
\begin{enumerate}
    \item \textbf{AMP (Oracle-$K$)}: assumes exact knowledge of the sparsity $\|\boldsymbol{x}\|_0$ for \emph{each} test vector.
    \item \textbf{AMP (Noisy-$K$)}: uses a perturbed sparsity estimate $\hat{K} = \|\x\|_0 + \delta$, where $\delta$ is an integer-valued, uniformly distributed random variable bounded as $\delta \in \{-\Delta, \dots, \Delta\}$.
    \item \textbf{AMP (Fixed-$K$)}: assumes a single global sparsity value $K$ across all test samples.
\end{enumerate}
All variants of AMP were executed using the code provided by the authors of \cite{tan2024approximate}. Since the output of the AMP method is not binary but a real-valued vector, we threshold the output element-wise using the threshold that maximizes the Success Rate on a validation set $\mathcal{D}_\text{val}^\text{AMP}$ of $n_\text{val}^\text{AMP}$ vectors generated using the same distribution as in Section \ref{sec:problemformulation}. They were tested on a dataset of $n_\text{test}^\text{AMP}$ vectors generated from the same distribution.

In contrast to AMP variants, the proposed MLP decoder is trained end-to-end on measurement data and does not receive the true sparsity or its estimate at inference time. Recall that as per the generative model in \eqref{eq:gen_K}, the \emph{average} $\ell_0$ norm of the generated sparse signals is $K$. However, their actual $\ell_0$ norm can differ significantly from $K$.

\begin{figure}[!t]
    \centering

    \begin{subfigure}[t]{0.32\textwidth}
        \centering
        \includegraphics[width=\linewidth]{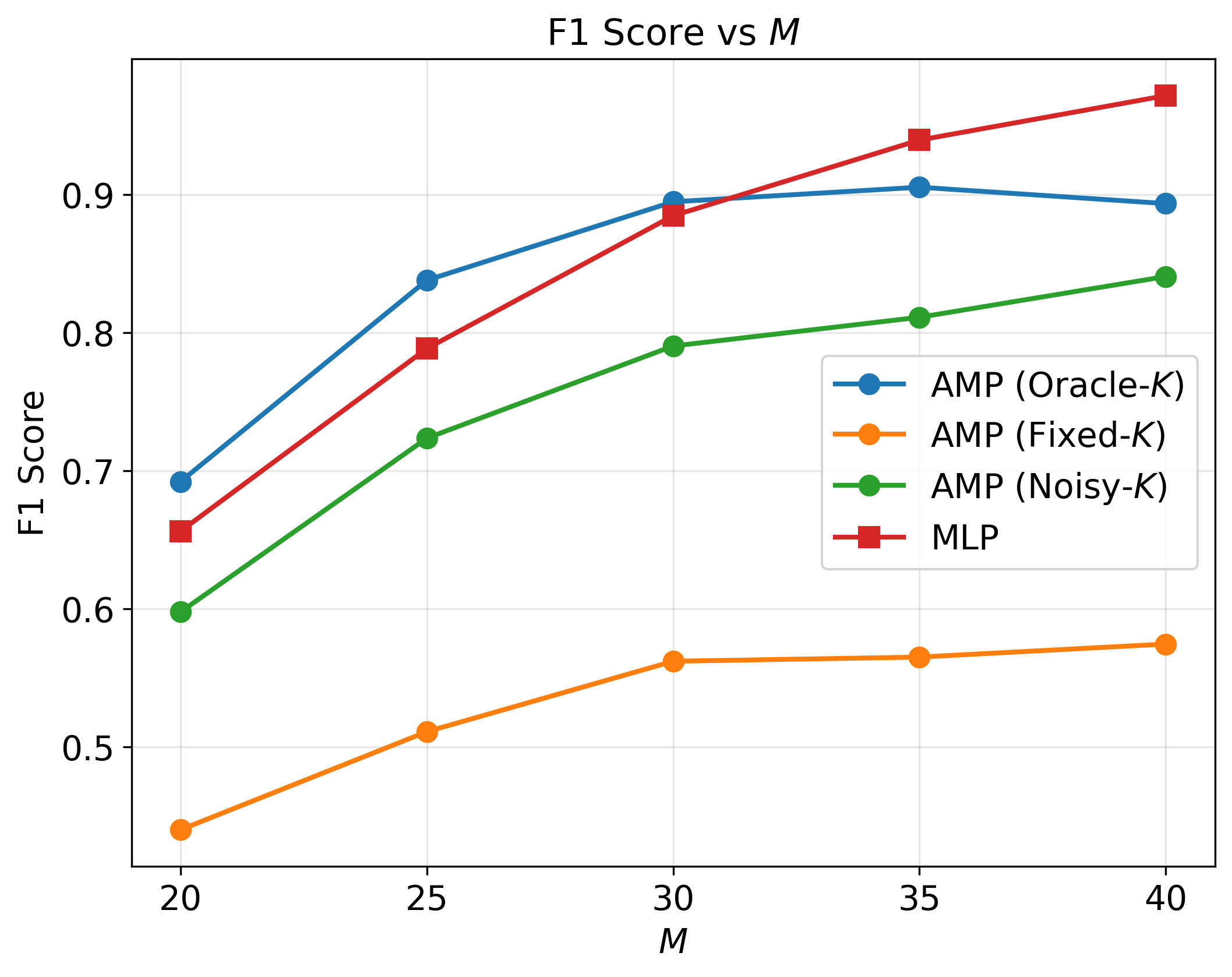}
        \label{fig:f1scorevsm}
    \end{subfigure}
    \hfill
    \begin{subfigure}[t]{0.32\textwidth}
        \centering
        \includegraphics[width=\linewidth]{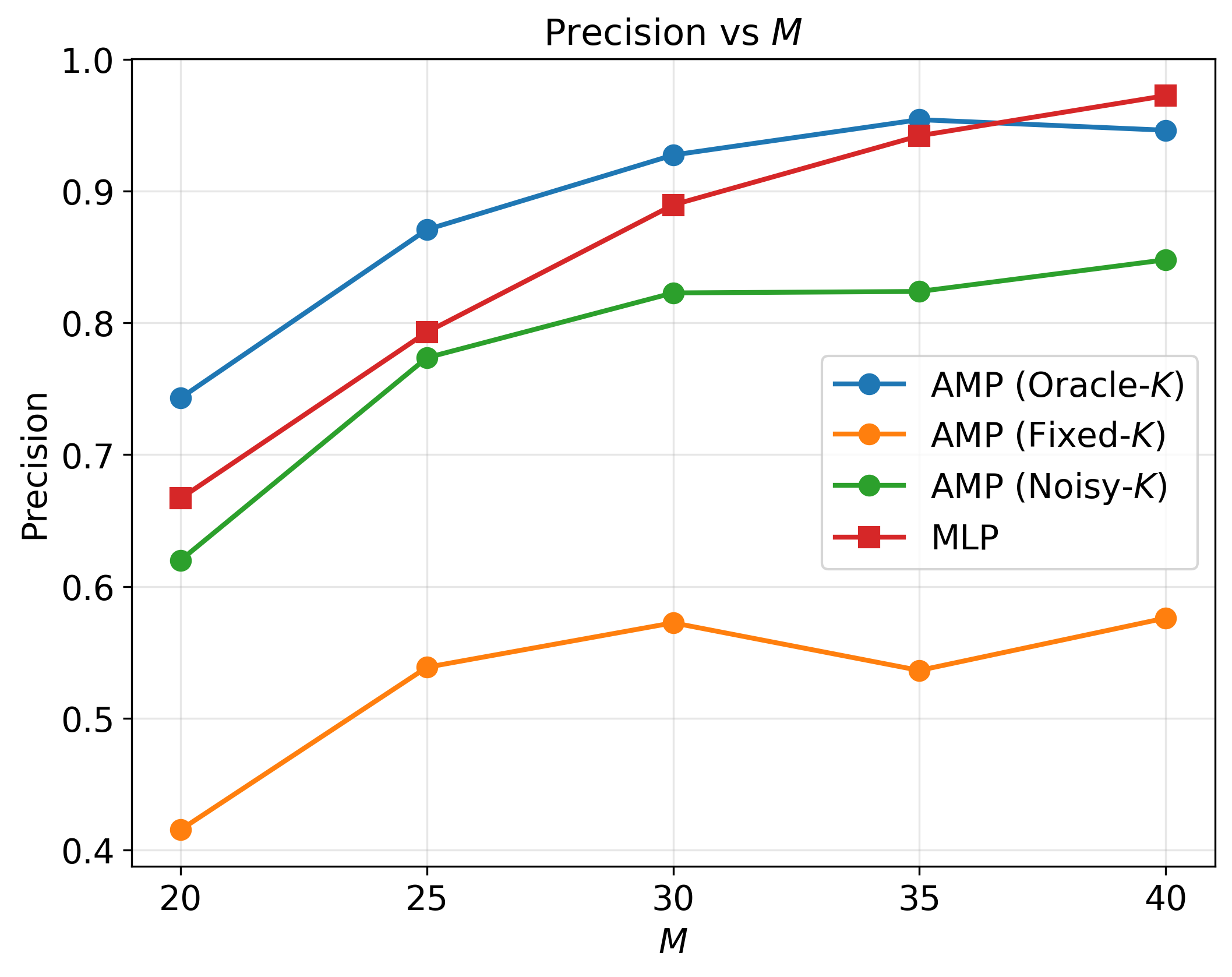}
        \label{fig:precisionvsm}
    \end{subfigure}
    \hfill
    \begin{subfigure}[t]{0.32\textwidth}
        \centering
        \includegraphics[width=\linewidth]{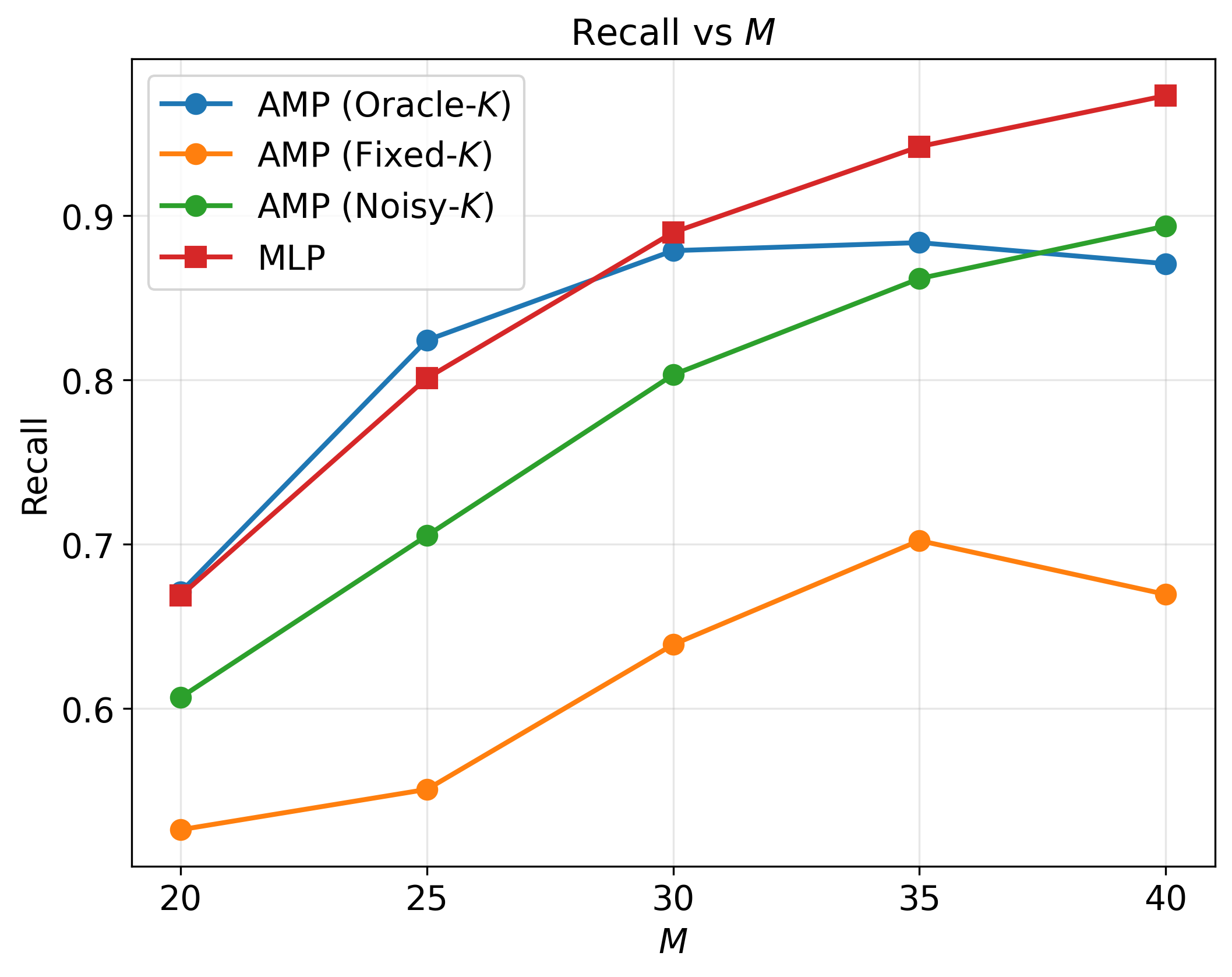}
        \label{fig:recallvsm}
    \end{subfigure}

    \vspace{0.2cm}

    \begin{minipage}{0.66\textwidth}
        \centering
        \begin{subfigure}[t]{0.48\textwidth}
            \centering
            \includegraphics[width=\linewidth]{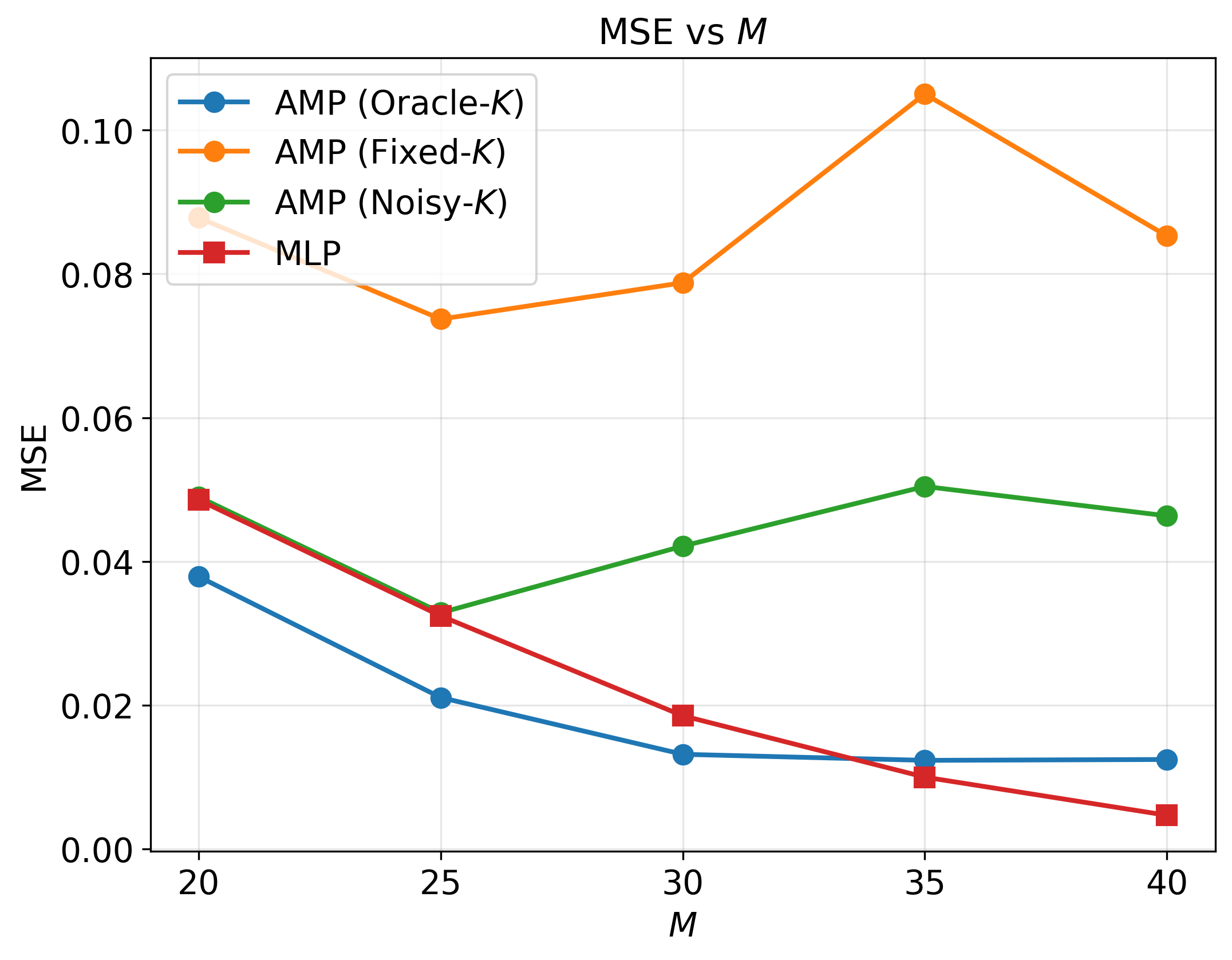}
            \label{fig:msevsm}
        \end{subfigure}
        \hfill
        \begin{subfigure}[t]{0.48\textwidth}
            \centering
            \includegraphics[width=\linewidth]{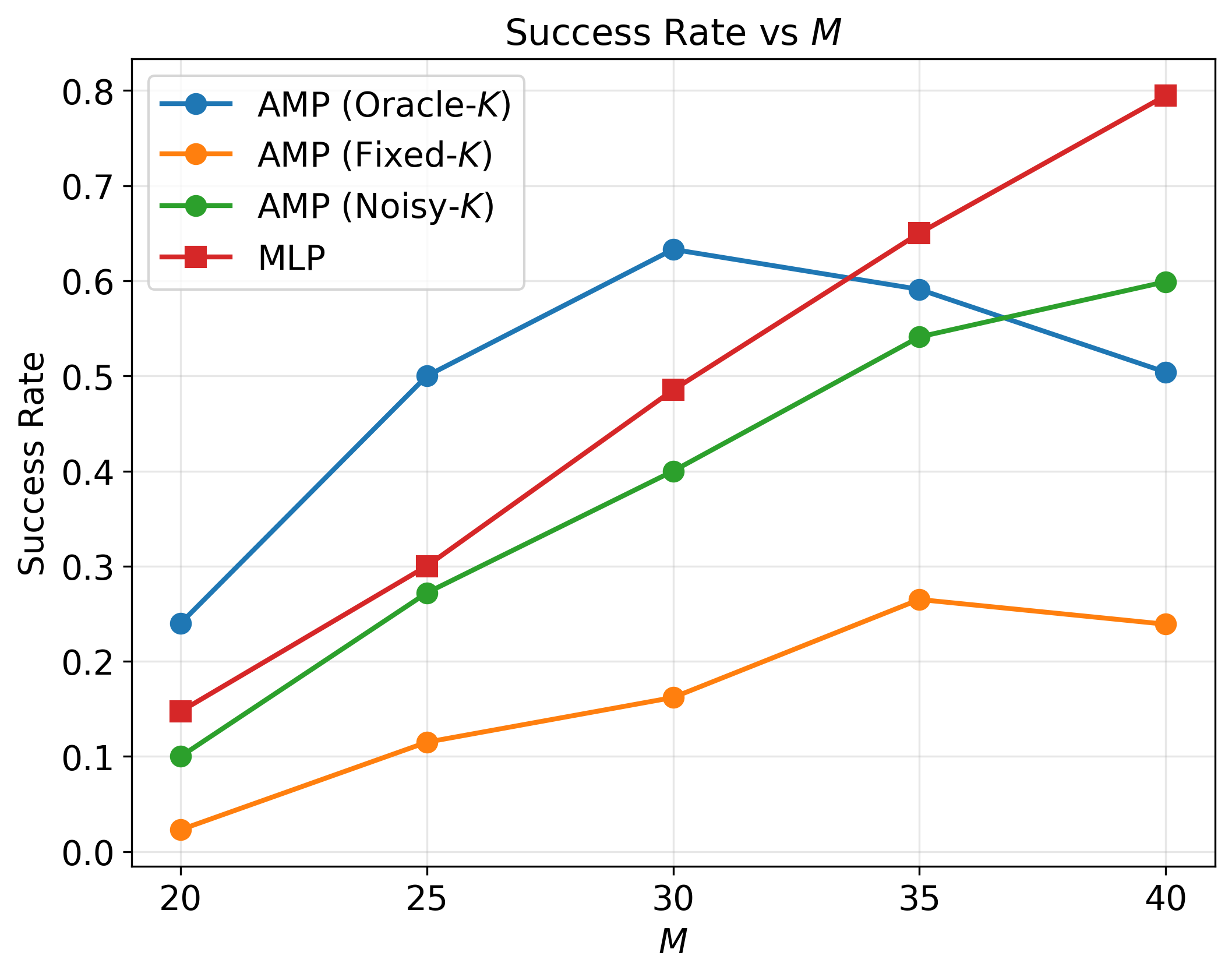}
            \label{fig:successratevsm}
        \end{subfigure}
    \end{minipage}

    \caption{
    Decoding performance versus the number of measurements $M$ for the proposed MLP decoder and AMP baselines. Each curve shows the average measure value as a function of $M$, with all other parameters fixed ($N = 100$, $K/N = 0.06$, $S/N = 0.1$, $D = 1$). The MLP achieves performance comparable to AMP (Oracle-$K$) without requiring knowledge of sparsity and remains robust across different measurement regimes.
    }
    \label{fig:metrics_vs_M}
\end{figure}

We evaluate all methods using five different measures, computed for each test vector individually and averaged over the entire test set:
\begin{gather*}
\text{Precision } (P) = \frac{TP}{TP + FP}, \quad
\text{Recall } (R) = \frac{TP}{TP + FN}, \quad
F_1 = \frac{2PR}{P + R}, \\
\text{SR} = \mathbf{1}\{\hat{\boldsymbol{x}}_{\text{bin}} = \boldsymbol{x}\}, \quad
\text{MSE} = \frac{1}{N}\|\hat{\boldsymbol{x}}_{\text{bin}} - \boldsymbol{x}\|_2^2,
\end{gather*}
where $TP$, $FP$, and $FN$ denote the numbers of true positives, false positives, and false negatives, respectively. SR denotes the Success Rate (fraction of samples with exact support recovery), and MSE quantifies the reconstruction error after thresholding.

We conduct two main experiments to compare the MLP decoder with the AMP baselines, using a primary setup of $N=100$ items and an average defect rate of $K/N=0.06$. For the AMP (Noisy-$K$) variant, we set the perturbation bound $\Delta=1$. The specific dataset sizes used for training, validation, and testing of all methods ($n_\text{train}^\text{MLP}$, $n_\text{test}^\text{MLP}$, $n_\text{val}^\text{AMP}$, etc.) are detailed in Appendix \ref{app:dataset_sizes}.

First, we evaluate performance as a function of the number of measurements $M$, varying it from 20 to 40. For this experiment, we fix the noise parameters at $S/N=0.1$ and $D=1$. As shown in Figure \ref{fig:metrics_vs_M}, the proposed MLP approach consistently outperforms both AMP (Noisy-$K$) and AMP (Fixed-$K$). Notably, the MLP's performance is comparable to, and in many regimes superior to, the AMP (Oracle-$K$) baseline, despite the MLP having no access to sparsity information.

\begin{figure}[t]
    \centering

    \begin{subfigure}[t]{0.32\textwidth}
        \centering
        \includegraphics[width=\linewidth]{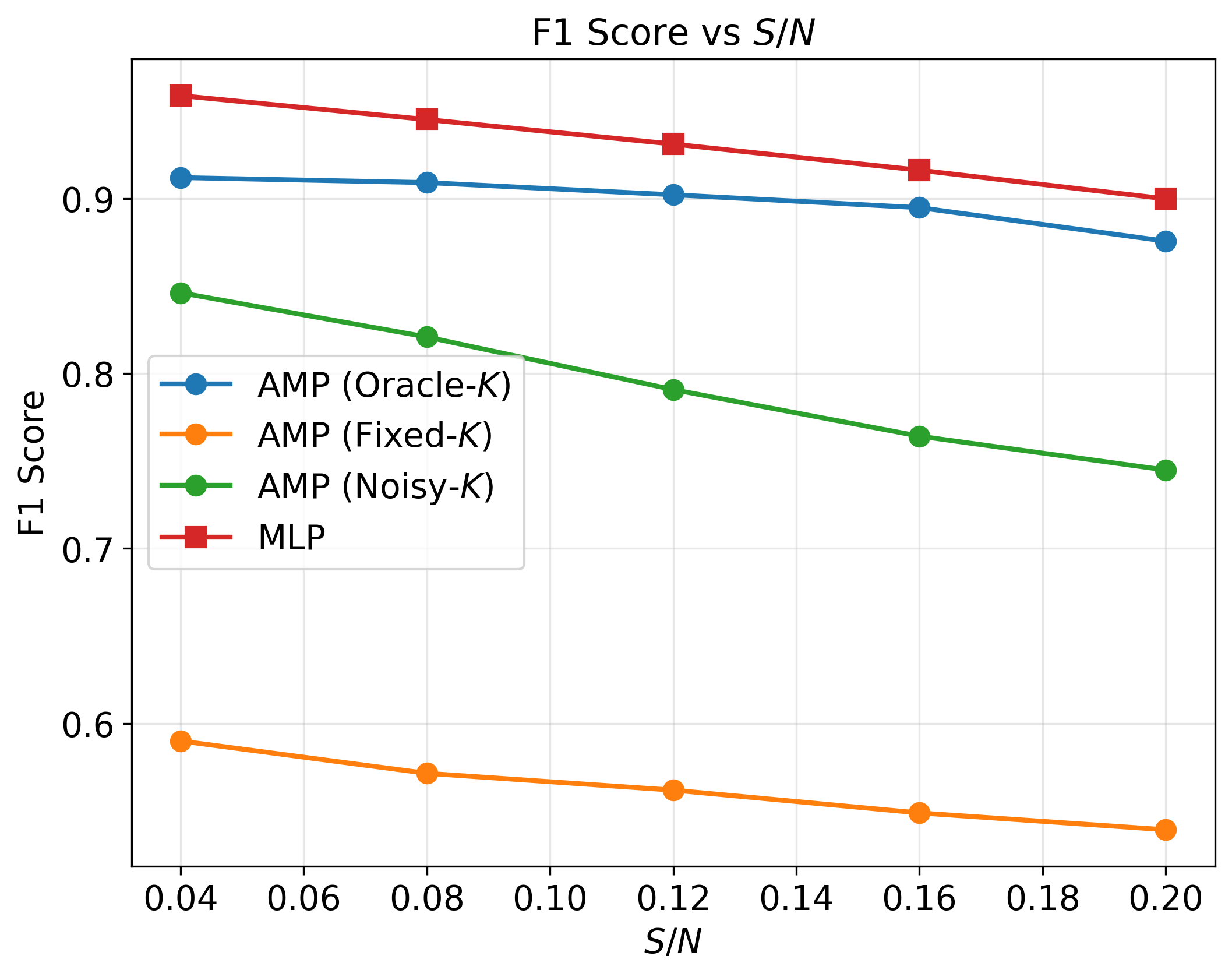}
        \label{fig:f1scorevsnoises}
    \end{subfigure}
    \hfill
    \begin{subfigure}[t]{0.32\textwidth}
        \centering
        \includegraphics[width=\linewidth]{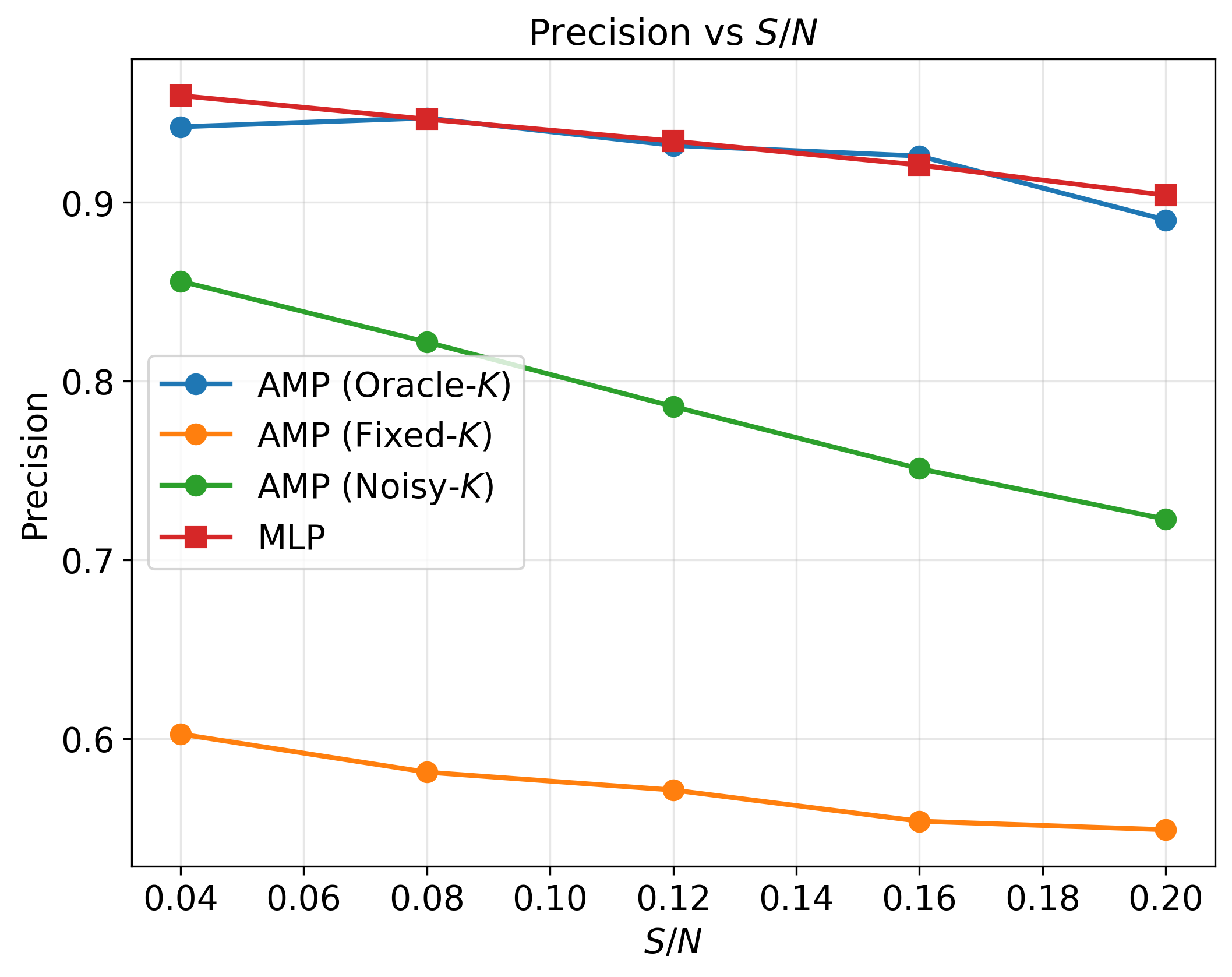}
        \label{fig:precisionvsnoises}
    \end{subfigure}
    \hfill
    \begin{subfigure}[t]{0.32\textwidth}
        \centering
        \includegraphics[width=\linewidth]{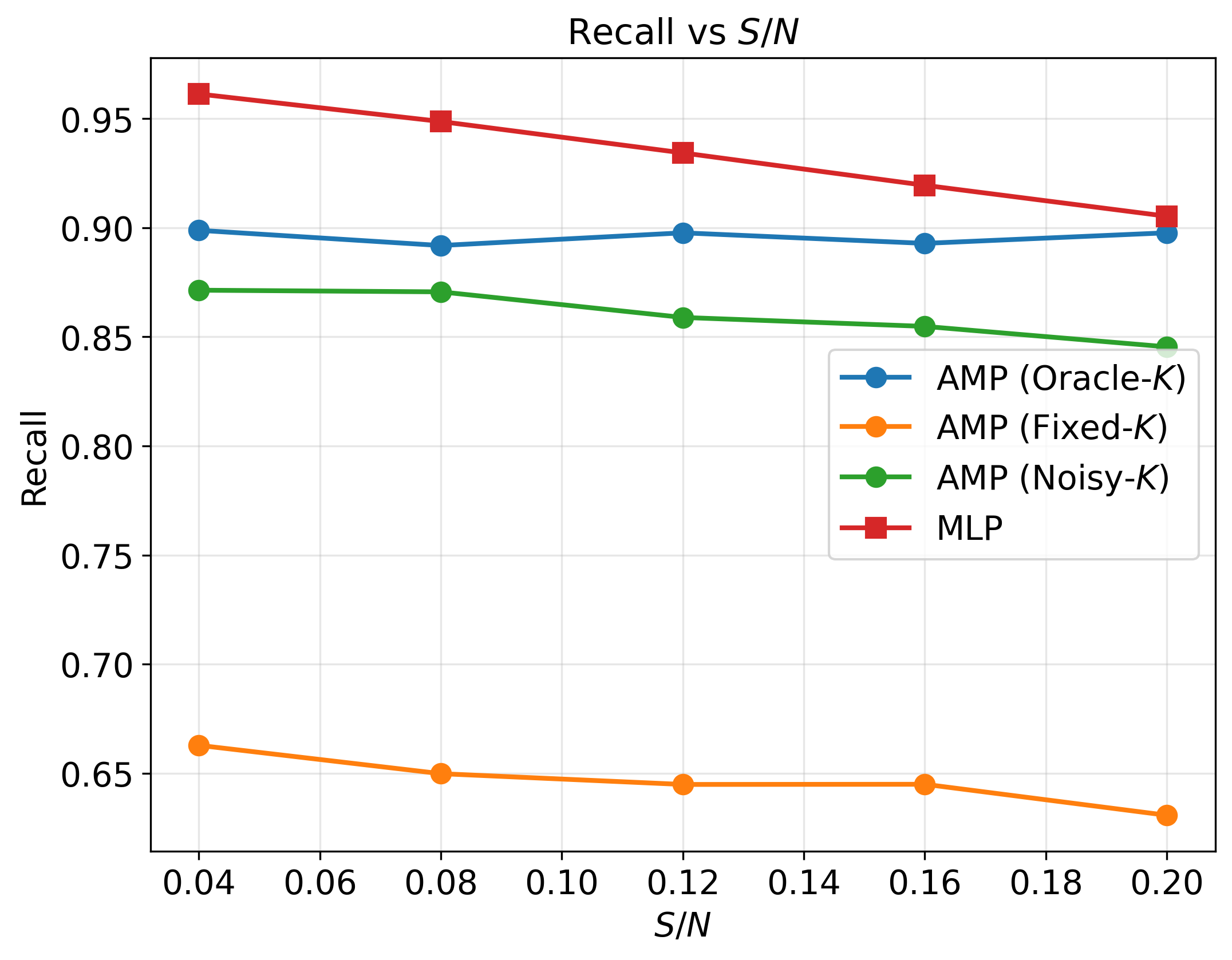}
        \label{fig:recallvsnoises}
    \end{subfigure}

    \vspace{0.2cm}

    \begin{minipage}{0.66\textwidth}
        \centering
        \begin{subfigure}[t]{0.48\textwidth}
            \centering
            \includegraphics[width=\linewidth]{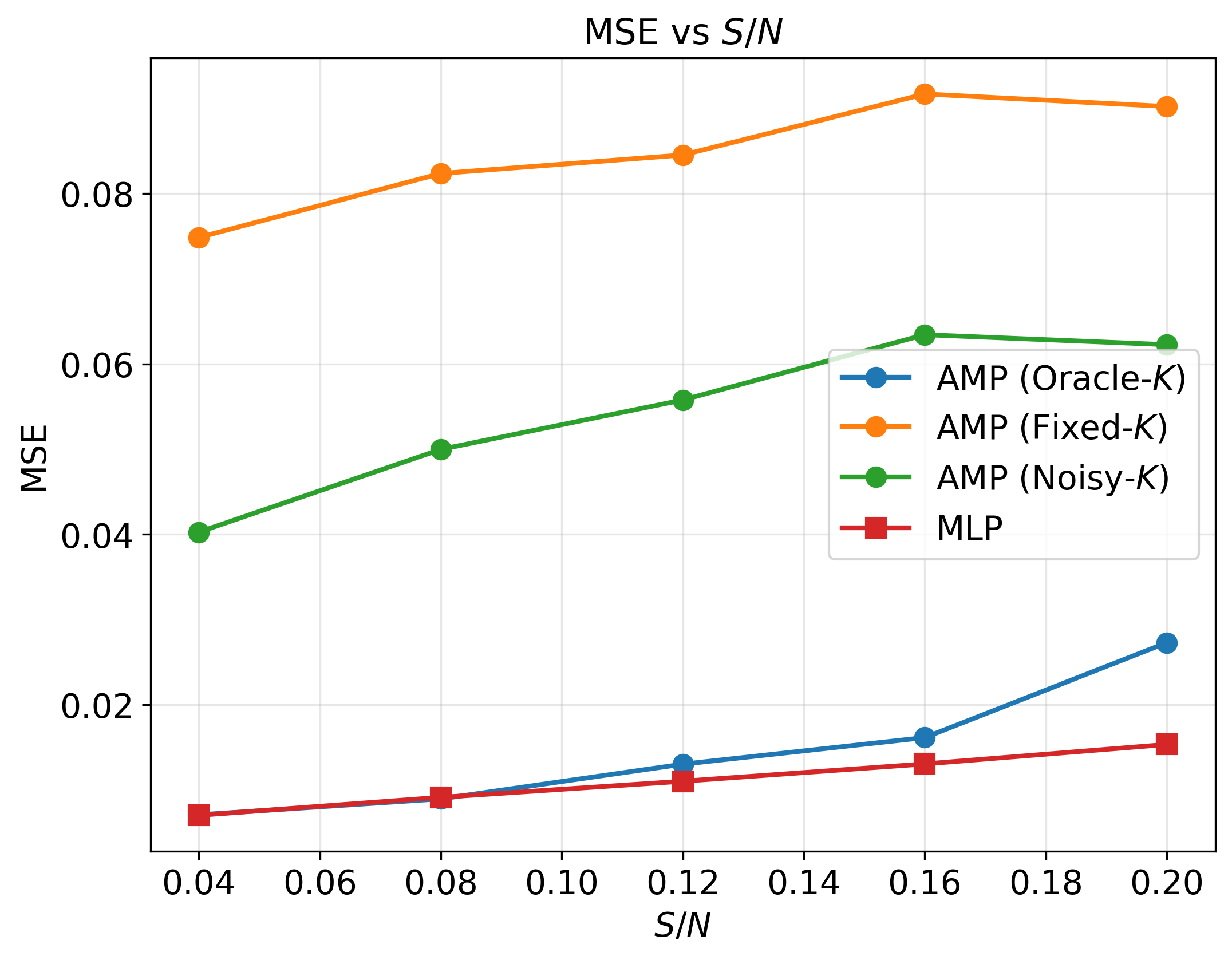}
            \label{fig:msevsnoises}
        \end{subfigure}
        \hfill
        \begin{subfigure}[t]{0.48\textwidth}
            \centering
            \includegraphics[width=\linewidth]{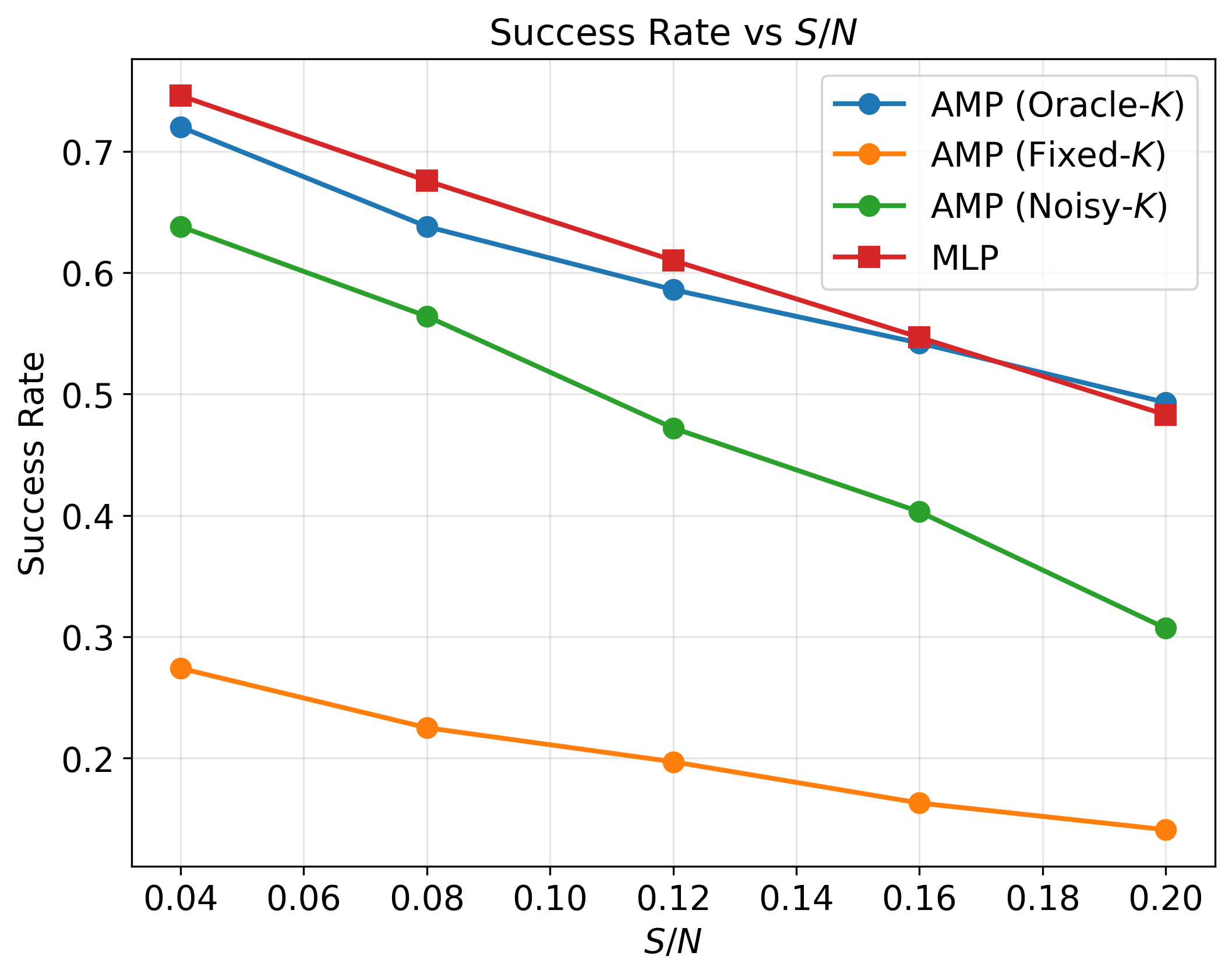}
            \label{fig:successratevsnoises}
        \end{subfigure}
    \end{minipage}

    \caption{
    Decoding performance versus noise sparsity $S$ for the proposed MLP decoder and AMP baselines. Each curve shows the average measure value as a function of $S$, with all other parameters fixed ($N = 100$, $K/N = 0.06$, $M = 35$, $D = 1$). The MLP matches or surpasses AMP (Oracle-$K$) across all tested noise levels, demonstrating strong robustness to sparse measurement noise.
    }
    \label{fig:metrics_vs_S}
\end{figure}

Second, we analyze robustness to noise by varying the noise sparsity $S/N$ from $0.04$ to $0.20$, while keeping other parameters fixed ($M=35$, $K/N=0.06$, $D=1$). Figure \ref{fig:metrics_vs_S} illustrates that the MLP decoder achieves performance on par with, or even superior to, the AMP (Oracle-$K$) method across all tested noise levels. This result highlights the MLP's strong robustness against sparse measurement noise.

\subsection{Effect of Network Complexity on Recovery and Verifiability}
\label{subsec:effectnetworkcomplexity}

To study how model capacity affects decoding and $\boldsymbol{A}$-recovery, we evaluate seven 
architectures with varying depth and width under identical training conditions. These span a wide 
range of expressive capacities, from a single linear layer (Level~1) to a deeper five-hidden-layer 
MLP (Level~7). The exact architectural specifications are provided in Appendix \ref{app:model_complexity}.

Table~\ref{tab:complexity_trend} summarizes the results for $N = 100$, $M = 35$, $K = 6$, $D = 1$, $S/N = 0.06$, and $T = 1000$, with all reported values averaged over five independent training runs, each using a different randomly generated pooling matrix $\boldsymbol{A}$. While increasing the nominal architectural complexity does not lead to strictly monotonic improvements in performance, a clear pattern emerges: models of moderate depth (Complexity Levels 3–5) achieve substantially lower $\boldsymbol{A}$-recovery error and higher decoding accuracy compared to both very small and very large architectures. This behavior is consistent with the fact that neural architectures with different widths, depths, and normalization layers do not define a single linear complexity scale—architectural choices interact with optimization dynamics in ways that can produce non-monotonic performance.

Despite this, the trend is unambiguous: architectures with sufficient but not excessive capacity learn the combinatorial structure governing QGT most effectively. In these models, the percentage of mismatched entries between $\boldsymbol{A}$ and $\boldsymbol{\hat{A}}$ drops sharply, and both the F1-Score and Success Rate improve accordingly.

\begin{table}[t]
\centering
\caption{Effect of network complexity on decoding and $\boldsymbol{A}$-recovery performance for $N = 100$, $M = 35$, $K = 6$, $D = 1$, and $S/N = 0.06$. Reported values are averaged over five random seeds corresponding to different $\boldsymbol{A}$ matrices and training runs. The ``Error (\%)'' column denotes the percentage of mismatched entries between $\boldsymbol{A}$ and $\boldsymbol{\hat{A}}$.}
\begin{tabular}{|l|c|c|c|}
\hline
\textbf{Model Complexity} & \textbf{Error (\%)} & \textbf{F1-Score} & \textbf{Success Rate (\%)} \\
\hline
Level 1 & 46.91 & 0.78 & 0.17 \\
Level 2 & 0.75 & 0.90 & 0.47 \\
Level 3 & 0.32 & 0.91 & 0.50 \\
Level 4 & 0.14 & 0.93 & 0.61 \\
Level 5 & 0.26 & 0.95 & 0.71 \\
Level 6 & 3.31 & 0.95 & 0.69 \\
Level 7 & 11.07 & 0.94 & 0.67 \\

\hline
\end{tabular}
\label{tab:complexity_trend}
\end{table}

\section{Conclusion}
In this work, we introduced a data-driven framework for Quantitative Group Testing (QGT) that achieves robust recovery under sparse noise without relying on explicit sparsity estimates during inference. By training on a representative signal distribution, our MLP decoder outperforms methods like Approximate Message Passing (AMP) in noisy regimes. Crucially, we demonstrated that our framework is \textbf{structurally verifiable}. Through Jacobian-based analysis, we showed that the trained network does not merely memorize training patterns but implicitly reconstructs the underlying pooling matrix $\boldsymbol{A}$, enabling it to reconstruct signals with support sets that are \emph{unseen} during training. This finding confirms that the model internalizes the true combinatorial relationships governing the problem, bridging the gap between black-box deep learning and model-based inverse problems. Future research will focus on extending this framework to the recovery of real-valued signals $\x$, a generalization with significant practical implications for applications like COVID-19 screening, where estimating viral loads is critical. Additionally, theoretical analysis of the proposed data-driven solver remains a direction for future work.

\bibliography{main}
\bibliographystyle{tmlr}

\appendix
\section{Appendix}

\subsection{Architectural and Training Details}
\label{app:arch_details}

The primary neural network architecture employed in this work (referred to as the MLP decoder) is a fully connected multi-layer perceptron. The network consists of an input layer of dimension $M$, two hidden layers, and an output layer of dimension $N$.

The specific architectural specifications are as follows:
\begin{itemize}
    \item \textbf{Input Layer:} Dimension $M$ (number of measurements).
    \item \textbf{Hidden Layers:} Two hidden layers, each containing 500 neurons.
    \item \textbf{Output Layer:} Dimension $N$ (number of items), corresponding to the real-valued score for each item.
    \item \textbf{Activation Function:} Leaky ReLU with a negative slope of $0.01$ is applied after each hidden layer.
    \item \textbf{Regularization:} Batch Normalization and Dropout with a probability of $p=0.1$ are applied after the activation of each hidden layer to prevent overfitting and improve generalization.
\end{itemize}

The model is trained using the \textbf{Adam} optimizer. To handle the class imbalance inherent in the sparse QGT problem, we utilize a balanced mean squared error loss function (as defined in \eqref{eq:balanced_mse}). Training incorporates early stopping based on the validation loss to determine the optimal number of epochs.

\subsection{Dataset Sizes}
\label{app:dataset_sizes}

The datasets used for training, validating, and testing the MLP, as well as for validating the AMP baselines, are summarized in Table~\ref{tab:dataset_sizes}. The training set is generated synthetically according to the probabilistic model described in Section \ref{sec:problemformulation}, ensuring that the network learns to generalize across the distribution of sparse signals and noise.

\begin{table}[h]
    \centering
    \caption{Summary of dataset sizes used in the experiments.}
    \label{tab:dataset_sizes}
    \begin{tabular}{lcc}
        \toprule
        \textbf{Dataset Split} & \textbf{Symbol} & \textbf{Number of Samples} \\
        \midrule
        MLP Training Set & $n_\text{train}^\text{MLP}$ & 119205 \\
        MLP Validation Set & $n_\text{val}^\text{MLP}$ & 14900 \\
        MLP Test Set & $n_\text{test}^\text{MLP}$ & 14900 \\
        \midrule
        AMP Validation Set & $n_\text{val}^\text{AMP}$ & 1000 \\ 
        AMP Test Set & $n_\text{test}^\text{AMP}$ & 1000 \\
        \bottomrule
    \end{tabular}
\end{table}

\subsection{Models of Different Complexities}
\label{app:model_complexity}

To analyze the impact of model capacity on recovery performance and structural verifiability (as discussed in Section \ref{subsec:effectnetworkcomplexity}), we evaluated seven distinct architectures, ranging from a simple linear mapping to deep multi-layer networks. 

All models share the same input dimension $M$ and output dimension $N$. With the exception of Level 1 (Linear), all architectures utilize the same regularization and activation scheme as the primary model: Batch Normalization, Leaky ReLU ($0.01$), and Dropout ($0.1$) after every hidden layer. The configurations are detailed in Table~\ref{tab:model_complexities}.

\begin{table}[h]
    \centering
    \caption{Architectural configurations for models of varying complexity (Level 1 to Level 7).}
    \label{tab:model_complexities}
    \begin{tabular}{cl}
        \toprule
        \textbf{Level} & \textbf{Hidden Layer Configuration} \\
        \midrule
        1 & None (Linear Map: $M \to N$) \\
        2 & [128] \\
        3 & [256] \\
        4 & [256, 256] \\
        5 & [500, 500]\\
        6 & [256, 512, 256] \\
        7 & [128, 256, 512, 256, 128] \\
        \bottomrule
    \end{tabular}
\end{table}

\end{document}

%% file: math_commands.tex
\usepackage{amsmath,amsfonts,bm}

\def\eqref#1{equation~\ref{#1}}

\def\1{\bm{1}}

\DeclareMathAlphabet{\mathsfit}{\encodingdefault}{\sfdefault}{m}{sl}
\SetMathAlphabet{\mathsfit}{bold}{\encodingdefault}{\sfdefault}{bx}{n}

